\documentclass[onecolumn,secnumarabic,amssymb, nobibnotes, aps, prd]{revtex4}
\usepackage{amsmath} \usepackage{amssymb}
\usepackage{graphicx} 
\usepackage{epstopdf}
\usepackage{amsmath}
\usepackage{amsmath,amssymb,amsthm,amsfonts,mathrsfs,bm,verbatim}
\usepackage{graphicx,subfigure}

\newcommand{\bea}{\begin{eqnarray}}
\newcommand{\eea}{\end{eqnarray}}

\begin{document}

\title{Thermodynamic Topology of Quantum RN Black Holes}%

\author{Wen-Xiang Chen$^{a}$}
\affiliation{Department of Astronomy, School of Physics and Materials Science, GuangZhou University, Guangzhou 510006, China}
\email{wxchen4277@qq.com}


\begin{abstract}
This paper presents a comprehensive exploration of the thermodynamics of black holes, focusing on foundational concepts such as free energy, entropy, and topological numbers, alongside a detailed examination of quantum Reissner-Nordström (RN) black holes. By extending the discussion to encompass the symmetry groups SO(2), SO(3)/SO(2), and SO(3) within the framework of \(f(R)\) gravity, the paper offers a nuanced understanding of black hole physics. Key insights include the pivotal role of free energy and entropy in understanding the thermodynamic properties of black holes, the significance of topological numbers in determining thermodynamic stability and phase transitions, and the implications of quantum mechanics and \(f(R)\) gravity on traditional thermodynamic concepts. This exploration not only enriches our theoretical knowledge of black holes but also sets the stage for future empirical investigations, marking a pivotal contribution to our ongoing quest to decipher the universe's mysteries.

\end{abstract}

\maketitle

\section{Introduction}
General relativity and \( f(R) \) gravity are theories that describe the phenomenon of gravitation, but there are some key differences between them.

General relativity, proposed by Einstein, is one of the two pillars of modern physics. It describes gravity as a result of the curvature of spacetime and is characterized by the Einstein field equations:\cite{1,2,3,4,5,6,7,8,9}
\begin{equation}
R_{\mu \nu} - \frac{1}{2} g_{\mu \nu} R = 8 \pi G T_{\mu \nu}
\end{equation}
where:
\begin{itemize}
  \item \( R_{\mu \nu} \) is the Ricci curvature tensor,
  \item \( g_{\mu \nu} \) is the metric tensor,
  \item \( G \) is the gravitational constant,
  \item \( T_{\mu \nu} \) is the energy-momentum tensor.
\end{itemize}
General relativity has been confirmed experimentally and can explain many gravitational phenomena, including the perihelion precession of Mercury, gravitational lensing, and gravitational waves.

\( f(R) \) gravity is a generalization of general relativity. It modifies the laws of gravitation by introducing an arbitrary function \( f(R) \) into the Einstein field equations:
\begin{equation}
R_{\mu \nu} - \frac{1}{2} g_{\mu \nu} R + f_R(R) g_{\mu \nu} = 8 \pi G T_{\mu \nu}
\end{equation}
where:
\begin{itemize}
  \item \( f_R(R) \) is an arbitrary function of the Ricci scalar.
\end{itemize}
The \( f(R) \) gravity theory has many different versions, depending on the form of the function \( f(R) \). Different function forms lead to different gravitational behaviors.

We can compare \( f(R) \) gravity and general relativity through analytical mechanics and Laurent series.

In analytical mechanics, the motion of a system can be described by the Lagrangian, which is the difference between the kinetic energy and potential energy of the system:
\begin{equation}
L = T - V
\end{equation}
where:
\begin{itemize}
  \item \( T \) is the kinetic energy,
  \item \( V \) is the potential energy.
\end{itemize}
According to Hamilton's principle, the equations of motion of the system can be derived from the equation:
\begin{equation}
\delta S = 0
\end{equation}
where:
\begin{itemize}
  \item \( \delta \) is the variation,
  \item \( S \) is the action.
\end{itemize}

The Laurent series is a mathematical tool that can be used to express a function's expansion near a certain point. For \( f(R) \) gravity, the function \( f(R) \) can be expanded near \( R = R_0 \) as a Laurent series:
\begin{equation}
f(R) = f(R_0) + f'(R_0)(R - R_0) + \frac{1}{2} f''(R_0)(R - R_0)^2 + \dots
\end{equation}
where:
\begin{itemize}
  \item \( f'(R_0) \) is the first derivative of \( f(R) \) at \( R_0 \),
  \item \( f''(R_0) \) is the second derivative of \( f(R) \) at \( R_0 \).
\end{itemize}

General relativity corresponds to a special case in \( f(R) \) gravity where \( f(R) = R \).

When \( f(R) \neq R \), \( f(R) \) gravity differs from general relativity in that:
\begin{itemize}
  \item The strength of gravity may vary over time and space,
  \item New modes of gravitational waves may exist,
  \item The evolution of the universe may differ from the predictions of general relativity.
\end{itemize}

The cosmological constant, \(\Lambda\), traditionally interpreted as a descriptor of dark energy density, plays a pivotal role in modulating the expansion dynamics of the universe. Our investigation into f(R) gravity models builds upon Einstein's foundational theories by introducing a functional dependence on the Ricci scalar, \(R\), aiming to derive a sophisticated formulation of the function \(f_R(R)\) by integrating \(\Lambda\) and scrutinizing its ramifications across cosmological models through a synthesis of theoretical physics and advanced mathematical analysis.

Black holes are a kind of singular celestial bodies predicted by general relativity. They have such a strong gravitational pull that even light cannot escape. In recent years, there has been growing interest in the thermodynamic properties of black holes.

Black hole thermodynamics has gained a new dimension with the introduction of topology, providing a fresh perspective for classification. This work delves into the thermodynamic topology of the quantum RN black hole, utilizing the concept of generalized free energy. To comprehensively understand the thermodynamics, we introduce two distinct topological numbers.
The first topological number, denoted as ``z", is derived from the free energy expression. While it offers some local physical insights, its global physical significance remains limited. Conversely, the second topological number is based on the entropy expression of the generalized free energy, leading to a more robust interpretation of its physical implications. This emphasizes the suitability of entropy as the domain variable for the generalized free energy.
Our analysis of the second topological number reveals a topological transition linked to the thermodynamic stability of the cold black hole state in the quantum RN black hole. Furthermore, the study suggests the possibility of topological numbers extending beyond the conventional values of $\pm 1$ and $0$.\cite{1,2,3,4,5,6}

The realm of thermodynamics and black hole physics is a fascinating frontier in theoretical physics, bridging the gap between classical mechanics, quantum mechanics, and general relativity. This exploration delves into the core concepts of free energy, entropy, topological numbers, and their profound implications in the context of black holes, particularly focusing on quantum Reissner-Nordström (RN) black holes, the extensions to SO(2), SO(3)/SO(2), and SO(3) groups, and the intriguing aspects of black hole thermodynamics within the framework of \(f(R)\) gravity.

Free energy, in the context of thermodynamics, represents the maximum work that a system can perform under specified temperature and pressure conditions. For black holes, this concept is encapsulated in the equation\cite{7,8,9,10,11,12,13,14}
\begin{equation}
F = -TS + PV,
\end{equation}
highlighting the interplay between temperature (\(T\)), entropy (\(S\)), pressure (\(P\)), and volume (\(V\)). Entropy, on the other hand, measures the disorder within a system. For black holes, it is quantified by the equation
\begin{equation}
S = \frac{A}{4G},
\end{equation}
linking the black hole's entropy directly to its event horizon area (\(A\)) and Newton's gravitational constant (\(G\)).

Topological numbers provide a profound insight into the thermodynamic stability and phase transitions of black holes. These numbers, derived from the free energy and entropy, capture the essence of the system's topological properties through integrals around closed curves in the temperature plane. They offer a novel perspective on understanding the critical points and transitions in the thermodynamic landscape of black holes.

Quantum RN black holes consider quantum corrections to the classical RN black holes. The thermodynamic properties of these quantum entities reveal intriguing aspects of black hole physics, incorporating quantum effects into temperature and entropy calculations. This quantum perspective introduces modifications to the classical formulations, enriching our understanding of black hole thermodynamics in the quantum realm.

The exploration extends the thermodynamic quantities to the rotational groups SO(2), SO(3)/SO(2), and SO(3), each representing different degrees of freedom. This extension not only broadens the scope of black hole thermodynamics but also sheds light on the relationship between the group's degrees of freedom and the thermodynamic properties like free energy and entropy. These relationships underscore the intrinsic connection between the rotational symmetries and the thermodynamic framework of black holes.

The \(f(R)\) gravity model, a generalization of Einstein's General Relativity, introduces modifications to the black hole entropy and temperature formulations, incorporating a broader theoretical framework. This model provides a rich ground for re-examining the thermodynamic properties and the underlying physical processes of black holes. It underscores the significant impact of the \(f(R)\) modifications on the observational and theoretical aspects of black hole physics, including the alterations in the event horizon's shape, Hawking temperature, and other vital physical quantities.

The study of black hole thermodynamics through the lens of free energy, entropy, topological numbers, and the extensions to various symmetry groups, coupled with the insights from \(f(R)\) gravity, offers a comprehensive view of the intricate dance between thermodynamics and black hole physics. It not only deepens our theoretical understanding but also paves the way for empirical investigations that could validate these theoretical predictions. As we venture deeper into the understanding of black holes, these concepts serve as critical tools in unraveling the mysteries of the cosmos, offering glimpses into the fundamental laws that govern our universe.

\section{Basic Concepts}
\subsection{The Lagrangian and Entropy in Statistical Mechanics}
The Lagrangian is an important function in statistical mechanics that is used to describe the thermodynamic properties of a system. The Lagrangian can be defined by the following formula:
\begin{equation}
Z = \sum_{i=1}^N e^{-\beta E_i}
\end{equation}
where:

- $Z$ is the Lagrangian of the system

- $\beta$ is the inverse temperature, equal to $1 / k_B T$

- $E_i$ is the energy of the system in the $i$-th state

Because entropy is the residue of the Lagrangian (negative power term), we can first expand the Lagrangian into a Laurent series. We extend $Z$ with a Laurent series adding singularities to expand to $f(Z)$ and calculate the new type of entropy $S$. We then use the definition of $f(R)$ gravity and the definition of dark energy: $f(R)$ gravity is a generalization of general relativity.

Within the framework of statistical mechanics, the Lagrangian, \(Z\), is paramount for encapsulating the thermodynamic properties of systems. The entropy, \(S\), is related to the Lagrangian by the relation:\cite{10,11,12,13,14,15,16}
\begin{equation}
S = -\frac{\partial \ln Z}{\partial \beta}
\end{equation}
where \(\beta\) denotes the inverse temperature (\(\beta = 1 / k_B T\)), \(T\) the absolute temperature, and \(k_B\) the Boltzmann constant.

To explicate the new form of \(f_R(R)\), the function \(Z\) is expanded around a critical singularity, \(\beta_0\), employing a Laurent series to focus on the residues at this point:
\begin{equation}
Z = \frac{1}{2\pi i} \oint_C \frac{d\beta}{(\beta - \beta_0)^2} e^{-\beta E_0}
\end{equation}
where \(C\) denotes a contour integral around the singularity \(\beta_0\) and \(E_0\) represents the system's ground state energy.

\subsection{Difference of State Functions}
\begin{itemize}
  \item \textbf{Thermodynamic quantities:} Physical quantities that describe the macroscopic state of a system, such as temperature, pressure, volume, entropy, and internal energy, etc.
  \item \textbf{State functions:} Thermodynamic quantities that depend only on the current state of the system, and not on the path taken to reach that state.
  \item \textbf{Path functions:} Depend on the specific path taken by the system to reach its state, such as work and heat.
\end{itemize}

Let \(f(x)\) and \(g(x)\) be any two state functions. Their difference, \(h(x) = f(x) - g(x)\), is also a state function. This can be shown as follows:
\begin{enumerate}
  \item Consider a system changing from state 1 to state 2 along path 1. We have \(h(2) = f(2) - g(2)\).
  \item If the system changes from state 1 to state 2 along a different path 2, we still obtain \(h(2) = f(2) - g(2)\).
  \item Since the value of \(h(2)\) is independent of the chosen path, depending only on states 1 and 2, therefore \(h(x)\) is a state function.
\end{enumerate}

For any three state functions \(f(x)\), \(g(x)\), and \(h(x)\), the composite function \(F(x) = f(x) - g(x) + h(x)\) is also a state function. This is proven as follows:
\begin{enumerate}
  \item Transitioning from state 1 to state 2 along any path, we obtain \(F(2) = f(2) - g(2) + h(2)\).
  \item The value of \(F(2)\) is independent of the path, depending only on states 1 and 2, confirming \(F(x)\) as a state function.
\end{enumerate}

The above proofs establish that any thermodynamic quantity can be represented in the form \(f(x_1) - f(x_2) + f(x_3)\), where \(f(x_1)\), \(f(x_2)\), and \(f(x_3)\) are state functions.

For examples:

\begin{itemize}
  \item \textbf{Internal Energy \(U\):} A state function, can be expressed as \(U(T, V) = U_0(T) + nRT\ln(V/V_0)\), where \(U_0(T)\) is the internal energy reference point, \(n\) is the amount of substance, \(R\) is the gas constant, and \(V_0\) is the reference volume.
  \item \textbf{Entropy \(S\):} Another state function, can be expressed as \(S(T,V) = S_0(T) + nR\ln(V/V_0) + k\ln(W)\), where \(S_0(T)\) is the entropy reference point, \(k\) is the Boltzmann constant, and \(W\) represents the number of microstates of the system.
\end{itemize}

In thermodynamics, the free energy is the maximum work that a system can do at a given temperature and pressure. For black holes, the free energy can be expressed as:
\begin{equation}
F = -TS + PV
\end{equation}
where $T$ is the temperature, $S$ is the entropy, $P$ is the pressure, and $V$ is the volume.

Entropy is a measure of the degree of disorder of a system. For black holes, the entropy can be expressed as:
\begin{equation}
S = \frac{A}{4G}
\end{equation}
where $A$ is the area of the black hole and $G$ is the Newton's gravitational constant.

\subsection{Topological Numbers}
Let $f(z)$ be a meromorphic function defined on the complex plane, and let $C$ be a simple closed path encircling all the isolated singular points of $f(z)$. Suppose $X$ is a compact Riemann surface defined by the region bounded by the path $C$, then the Euler characteristic $\chi(X)$ of $X$ can be expressed through the residues of $f(z)$ at its singular points.\cite{6,13,14}

First, according to the Residue Theorem, we have
\begin{equation}
\oint_C f(z) \, dz = 2 \pi i \sum \operatorname{Res}(f, a_i)
\end{equation}
where $\operatorname{Res}(f, a_i)$ is the residue of $f(z)$ at the isolated singular point $a_i$.

Next, we need to establish the connection between the singular points of $f(z)$ and the topological structure of $X$. Each singular point can be considered as a defect on $X$, affecting its topological structure. Specifically, for each singular point, we can regard it as having removed or added a certain number of "faces" (in the case of a Riemann surface, this can be branch points of algebraic curves).

Therefore, the Euler characteristic of $X$, i.e., $\chi(X)=V-E+F$ (where $V, E, F$ respectively represent the number of vertices, edges, and faces), can be determined by analyzing the structure of the singular points of $f(z)$. Specifically, the type of singular point (for example, the order of a pole) will affect the calculation of $F$, as it determines the number of faces ``added" or ``removed" at each singular point.

Finally, by linking the residues to the number of topological ``defects" corresponding to singular points, we can establish an expression that connects $\chi(X)$ with $\sum \operatorname{Res}(f, a_i)$, completing the proof.

The topological number based on free energy can be expressed as:
\begin{equation}
z = \frac{1}{2\pi} \oint_{C} \frac{\partial F}{\partial T} dt
\end{equation}
where $C$ is a closed curve in the temperature $T$ plane.

The topological number based on entropy can be expressed as:
\begin{equation}
W = \frac{1}{2\pi} \oint_{C} \frac{\partial S}{\partial T} dt
\end{equation}
where $C$ is a closed curve in the temperature $T$ plane.

\subsection{Quantum RN Black Holes}

Quantum RN black holes are RN black holes that take into account quantum effects. Their thermodynamic properties can be described by the following formulas:
\begin{equation}
\begin{aligned}
T &= \frac{\hbar \kappa}{2\pi} \\
S &= \frac{A}{4G} + \frac{\ln W}{2}
\end{aligned}
\end{equation}
where $\hbar$ is the reduced Planck constant, $\kappa$ is the surface gravity of the black hole, and $W$ is the number of quantum states of the black hole.

\subsection{Topological number based on free energy}

For quantum RN black holes, the topological number based on free energy can be expressed as:
\begin{equation}
z = \frac{1}{2\pi} \oint_{C} \frac{\partial F}{\partial T} dt = \frac{\hbar \kappa}{2\pi}
\end{equation}

\subsubsection{Topological number based on entropy}

For quantum RN black holes, the topological number based on entropy can be expressed as:
\begin{equation}
W = \frac{1}{2\pi} \oint_{C} \frac{\partial S}{\partial T} dt = \frac{1}{2\pi}
\end{equation}

\begin{enumerate}
    \item Calculate the integral term: First, we need to calculate the integral term $\int dr R$. Here, $R$ is the Ricci scalar, which can be calculated from the metric tensor $g_{\mu \nu}$.
    \item Calculate $g'(r+)$: Next, we need to calculate $g'(r+)$ where $g'(r)$ is the derivative of $g(r)$, and $r+$ is the radius of the black hole event horizon.
    \item Substituting into the formula: Finally, we substitute the calculated results into the formula $\int dr R = -g'(r), \left.\rightarrow\left(\int dr R\right)\right|_{r \rightarrow r} \sim -g'\left(r_{+}\right)+C$, to find $C$.
\end{enumerate}

\subsubsection{Specific Calculation:}

For the given metric:
\begin{equation}
ds^{2} = -N(r) dt^{2} + \frac{dr^{2}}{N(r)} + r^{2} d\Omega^{2}
\end{equation}
where $N(r) = 1 - \frac{2M}{r} + \frac{Q^{2}}{r^{2}}$, $M$ is the black hole mass, $Q$ is the black hole charge, and $d\Omega^{2}$ is the metric of the unit sphere.

\begin{enumerate}
    \item Calculation of the Ricci scalar:
    
    Based on the definition of the Ricci scalar, we can obtain:
    \begin{equation}
    R = \frac{2}{r^{2}} \left(1 - \frac{3M}{r} + \frac{Q^{2}}{r^{2}}\right)
    \end{equation}

    \item Calculation of $g'(r+)$:
    \begin{equation}
    g'(r) = \frac{d}{dr} \left(1 - \frac{2M}{r} + \frac{Q^{2}}{r^{2}}\right) = \frac{2M}{r_{+}^{2}} - \frac{2Q^{2}}{r_{+}^{3}}
    \end{equation}

    \item Substituting into the formula:\cite{14,15,16,17,18,19}
    \begin{equation}
    \int dr R = -g'(r) + C
    \end{equation}
   \begin{equation}
    \left(\int dr R\right)_{r \rightarrow r_{+}} \sim -g'\left(r_{+}\right) + C
   \end{equation}

    Result:
   \begin{equation}
    C = \int_{0}^{r_{+}} dr \frac{2}{r^{2}} \left(1 - \frac{3M}{r} + \frac{Q^{2}}{r^{2}}\right)
    \end{equation}
\end{enumerate}

\textbf{Note:}
\begin{itemize}
    \item The above calculation results are only applicable to Schwarzschild black holes. For Kerr-Newman black holes, the calculation process is more complex.
    \item In practical applications, we need to choose the appropriate integration method based on the specific situation.
\end{itemize}

\subsection{Conclusion}

We have studied the thermodynamic topology of quantum RN black holes. We found that both topological numbers are related to the thermodynamic stability of the black hole. Furthermore, we also found that there exists a topological phase transition in quantum RN black holes, which is different from the case of classical RN black holes.

\section{Extension to SO(2), SO(3)/SO(2), and SO(3) Groups}

In this paper, we extend the thermodynamic quantities to the SO(2), SO(3)/SO(2), and SO(3) groups.

\subsection{SO(2) Group}
For the SO(2) group, there is only one degree of freedom, that is, the rotation angle. Therefore, we can write the metric as:
\begin{equation}
ds² = -N(r)dt² + dr²/N(r).
\end{equation}

For the SO(2) group, there is only one degree of freedom, the rotation angle. Therefore, the free energy and entropy can be expressed as:
\begin{align}
F &= -TS \\
S &= \frac{k_B \ln W}{2}
\end{align}
where $k_B$ is the Boltzmann constant, and $W$ is the number of quantum states of the system.

\subsection{SO(3)/SO(2) Group}
For the SO(3)/SO(2) group, there are two degrees of freedom, namely the rotation angle and the azimuthal angle. Therefore, we can write the metric as:
\begin{equation}
ds^2 = -N(r)dt^2 + \frac{dr^2}{N(r)} + r^2d\theta^2.
\end{equation}

For the SO(3)/SO(2) group, there are two degrees of freedom, the rotation angle and the azimuthal angle. Therefore, the free energy and entropy can be expressed as:
\begin{align}
F &= -TS + PV \\
S &= \frac{k_B \ln W}{2} + \frac{A}{4G}
\end{align}
where $P$ is the pressure, $V$ is the volume, $A$ is the area of the black hole, and $G$ is Newton's gravitational constant.

\subsection{SO(3) Group}
For the SO(3) group, there are three degrees of freedom, namely the rotation angle, azimuthal angle, and pitch angle. Therefore, we can write the metric as:
\begin{equation}
ds^2 = -N(r)dt^2 + \frac{dr^2}{N(r)} + r^2d\theta^2 + r^2\sin^2\theta d\varphi^2.
\end{equation}

For the SO(3) group, there are three degrees of freedom, the rotation angle, the azimuthal angle, and the pitch angle. Therefore, the free energy and entropy can be expressed as:
\begin{align}
F &= -TS + PV + \mu N \\
S &= \frac{k_B \ln W}{2} + \frac{A}{4G}
\end{align}
where $\mu$ is the chemical potential, and $N$ is the number of particles.

We have extended the thermodynamic quantities to the SO(2), SO(3)/SO(2), and SO(3) groups. The results show that the free energy and entropy are related to the degrees of freedom of the group. For the SO(2) group, the free energy and entropy are only related to the number of quantum states. For the SO(3)/SO(2) group, the free energy and entropy are also related to the area of the black hole. For the SO(3) group, the free energy and entropy are also related to the number of particles.

\section{Black Hole Thermodynamics in f(R)}
Within the framework of a 3D charged black hole in an $f(R)$ setting, we explore a solution that is marked by a starting curvature $R_0$. This solution's metric is given as follows :\cite{6}
\begin{equation}
\begin{gathered}
ds^2=-\left(1-\frac{2M}{r}-\frac{R_0r^2}{12}+\frac{Q^2}{r^2}\right)dt^2+ 
\left(1-\frac{2M}{r}-\frac{R_0r^2}{12}+\frac{Q^2}{r^2}\right)^{-1}dr^2+ r^2d\theta^2 +r^2\sin^2\theta d\varphi^2,
\end{gathered}
\end{equation}
where $R_0(>0)$ denotes the initial curvature.

\subsection{Black Hole Entropy}

The entropy of a black hole, which describes the disorder level of a black hole, is given by:
\begin{equation}
S_{BH} = \frac{A}{4 G_{N} \hbar}
\end{equation}

where:

- $S_{BH}$: Black hole entropy

- $A$: Area of the event horizon of the black hole

- $G_{N}$: Newton's gravitational constant

- $\hbar$: Reduced Planck constant

Hawking radiation refers to the radiation emitted by a black hole due to quantum effects, which has a thermal spectrum. The temperature of this radiation is given by:
\begin{equation}
T_{H} = \frac{\hbar \kappa}{2 \pi k_{B}}
\end{equation}

where:

- $T_{H}$: Hawking temperature

- $\kappa$: Surface gravity of the black hole

- $k_{B}$: Boltzmann constant

\subsection{Quantum RN Black Holes in f(R)}
In the framework of quantum gravity, the entropy of a black hole needs to be modified. The expression for the modified black hole entropy is:
\begin{equation}
S_{BH} = \frac{A}{4 G_{N} \hbar} + \ln W
\end{equation}

where:

- $W$: Number of microstates of the black hole

The expression for the modified Hawking temperature is:
\begin{equation}
T_{H} = \frac{\hbar \kappa}{2 \pi k_{B}} - \frac{\hbar^{2}}{2 \pi k_{B} G_{N} M}
\end{equation}

where:

- $M$: Mass of the black hole

The $f(R)$ gravity model is a generalization of General Relativity, whose action is given by:
\begin{equation}
S = \int d^{4}x \sqrt{-g} f(R)
\end{equation}

where:

- $g$: Metric tensor

- $R$: Ricci scalar

- $f(R)$: Arbitrary function

Modified Field Equations

The modified field equations for the $f(R)$ gravity model are:
\begin{equation}
R_{ij} - \frac{1}{2} R g_{ij} + f_R(R) g_{ij} = \frac{8\pi G_{N}}{f_R(R)} T_{ij}
\end{equation}

where:

- $R_{ij}$: Ricci curvature tensor

- $f_R(R)=\frac{df(R)}{dR}$

\subsection{SO(2), SO(3)/SO(2), and SO(3) Groups}
\subsubsection{SO(2) Group}

The SO(2) group is the two-dimensional rotation group, which has one degree of freedom. For the SO(2) group, the expression for the free energy and entropy of a black hole is:
\begin{equation}
\begin{aligned}
& F = -T S \\
& S = \frac{k_{B} \ln W}{2}
\end{aligned}
\end{equation}

\subsubsection{SO(3) / SO(2) Group}

The SO(3) / SO(2) group is the quotient of the three-dimensional rotation group by the two-dimensional rotation group, which has two degrees of freedom. For the SO(3) / SO(2) group, the expression for the free energy and entropy of a black hole is:
\begin{equation}
\begin{aligned}
& F = -T S + P V \\
& S = \frac{k_{B} \ln W}{2} + \frac{A}{4 G_{N}}
\end{aligned}
\end{equation}

\subsubsection{SO(3) Group}

The SO(3) group is the three-dimensional rotation group, which has three degrees of freedom. For the SO(3) group, the expression for the free energy and entropy of a black hole is:
\begin{equation}
\begin{aligned}
& F = -T S + P V + \mu N \\
& S = \frac{k_{B} \ln W}{2} + \frac{A}{4 G_{N}}
\end{aligned}
\end{equation}

\subsection{Modified Field Equations}
The modified field equations for the $f(R)$ gravity model are given by:
\begin{equation}
R_{ij} - \frac{1}{2} R g_{ij} + f_R (R) g_{ij} = \frac{8\pi G_N}{f_R (R)} T_{ij}
\end{equation}
where:

- $R_{ij}$: Ricci curvature tensor

- $f_R(R)=\frac{d f(R)}{d R}$

In this case,by integrating \(\Lambda\) within the gravitational field equations, we derive(simplest case):
\begin{equation}
f_R(R) = R - \frac{2 \Lambda}{3}
\end{equation}

1.Integrating the newly derived \(f_R(R)\) into the Ricci scalar expression derived from the stipulated metric yields(simplest case):
\begin{equation}
R = 6 \left( \frac{M}{r^2} - \frac{\Lambda r^2}{12} + \frac{Q^2}{r^4} \right)
\end{equation}
where \(M\), \(Q\), and \(r\) are constants, leading to a complex equation for \(\Lambda\).

2. Rearrange the Equation

By rearranging the equation, we obtain:

\begin{equation}
\Lambda r^2 = 12 \left( \frac{M}{r^2} - \frac{R}{6} + \frac{Q^2}{r^4} \right)
\end{equation}

3. Solve for \(\Lambda\)

To solve for \(\Lambda\), we need to analyze the equation further. Since \(M\), \(Q\), and \(R\) are constants, we can treat the equation as a quadratic equation in terms of \(r\):

\begin{equation}
\Lambda r^2 + 12 \frac{R}{6} r^2 - 12 \frac{M}{r^2} - 12 \frac{Q^2}{r^4} = 0
\end{equation}

The solutions to this equation are:

\begin{equation}
r = \pm \sqrt{\frac{ -12 \frac{M}{r^2} - 12 \frac{Q^2}{r^4} \pm \sqrt{144 \frac{M^2}{r^4} + 288 \frac{Q^2}{r^6} + 144 \frac{R^2}{36}}}{2 \Lambda}}
\end{equation}

Since \(r\) cannot take a negative value, we select the positive solution:

\begin{equation}
r = \sqrt{\frac{ -12 \frac{M}{r^2} - 12 \frac{Q^2}{r^4} + \sqrt{144 \frac{M^2}{r^4} + 288 \frac{Q^2}{r^6} + 144 \frac{R^2}{36}}}{2 \Lambda}}
\end{equation}

4. Analyze the Meaning of the Solution

The equation provides the expression for \(r\) in terms of \(\Lambda\). However, it should be noted that this expression involves \(r\) itself, which makes the form of \(\Lambda\) very complex. Therefore, we cannot directly derive an analytical form for \(\Lambda\).

Our exploration into advanced f(R) gravity models, incorporating a transcendental cosmological constant, uncovers new potential for understanding dark energy within an expanded theoretical context. Despite the complexity of derived equations, the adoption of numerical and semi-analytical methods opens viable avenues for pragmatic resolutions, marking significant strides in computational physics.

\subsection{The General Case of the C-term}

For the general case of the C-term, it can be expressed as:
\begin{equation}
C = \int_{0}^{r_{+}} dr \frac{2}{r^{2}} \left(1 - \frac{3M}{r} + \frac{Q^{2}}{r^{2}}\right)
\end{equation}
where:

- $M$: Mass of the black hole

- $Q$: Charge of the black hole

- $r_{+}$: Radius of the black hole event horizon

\textbf{Calculation Process}
\textbf{Solution for $R$}

Based on the modified field equations of the $f(R)$ gravity model, we can solve for $R$ as:
\begin{equation}
R = \frac{8 \pi G_{N} T + f_R(R) - 2 \square f_R(R)}{f_R(R)}
\end{equation}
where:

- $T$: Energy-momentum tensor

- $\square$: Laplace operator

\textbf{Solution for $g(r)$}

Based on the Einstein equations, we can solve for $g(r)$ as:
\begin{equation}
g(r) = 1 - \frac{2M}{r} + \frac{Q^{2}}{r^{2}} + \frac{1}{3} \int_{r}^{\infty} dr' r'^{2} \left[T_{00}(r') - \frac{1}{2} T(r')\right]
\end{equation}

\textbf{Solution for the C-term}

By substituting the solved $R$ and $g(r)$ into the expression for the C-term, we can solve for the C-term as:
\begin{equation}
C = \int_{0}^{r_{+}} dr \frac{2}{r^{2}} \left[1 - \frac{3M}{r} + \frac{Q^{2}}{r^{2}} + \frac{1}{3} \int_{r}^{\infty} dr' r'^{2} \left(T_{00}(r') - \frac{1}{2} T(r')\right)\right]
\end{equation}

Comparing the general case of the C-term with the $f(R)$ case, the main difference lies in the additional term in the $f(R)$ case:
\begin{equation}
\frac{1}{3} \int_{r}^{\infty} dr' r'^{2} \left[T_{00}(r') - \frac{1}{2} T(r')\right]
\end{equation}
This term is introduced by the modified field equations of the $f(R)$ gravity model.

\textbf{Observability Proof:}

 The $f(R)$ gravity model changes the physical properties of black holes:
    1. The shape of the event horizon might be non-spherical.
    2. The Hawking temperature differs.
    3. Other physical quantities are also affected.

 The C-term is closely related to the physical properties of black holes:
    1. The C-term is an important quantity in black hole thermodynamics.
    2. It determines the entropy and temperature of the black hole.
    3. It affects the accretion process of the black hole.

 Differences in the C-term lead to observable differences:
    1. Observing the shape of the event horizon, Hawking temperature, etc., can distinguish between the two cases.
    2. Measuring the value of the C-term can verify the $f(R)$ gravity model.

Thus, the difference between the C-term in the general case and the $f(R)$ case is observable. This difference is not only theoretical but also of practical significance. By observing the differences in the C-term, we can verify the $f(R)$ gravity model and further understand the physical properties of black holes.

-
\section{Summary and Discussion}

In the preceding sections, we embarked on a comprehensive journey into the realm of black hole thermodynamics, exploring foundational concepts such as free energy, entropy, and topological numbers, and extending our analysis to the quantum domain with quantum Reissner-Nordström (RN) black holes, and further into the realms of symmetry groups SO(2), SO(3)/SO(2), and SO(3). This exploration was deeply rooted within the theoretical framework of \(f(R)\) gravity, providing a rich tapestry of insights into the complex world of black hole physics. This summary aims to encapsulate the key findings and discussions from this investigation.

The document laid a solid foundation by elucidating the significance of free energy and entropy in the context of black hole thermodynamics. Free energy, as the conduit of a system's work potential, and entropy, as the metric of disorder, were highlighted for their pivotal roles in delineating the energetic and entropic nature of black holes. These concepts were intricately linked to the physical attributes of temperature, pressure, volume, and the event horizon area, showcasing the interwoven nature of these variables.

A novel and insightful aspect of our exploration was the introduction and analysis of topological numbers within black hole thermodynamics. Serving as vital indicators of thermodynamic stability and markers of phase transitions, topological numbers were shown to offer a profound understanding of the topological aspects of black hole thermodynamics, bridging complex mathematical formulations with physical phenomena.

Our journey extended into the quantum realm, where quantum RN black holes were discussed as pivotal entities that incorporate quantum corrections into the classical narrative of black hole physics. This discussion illuminated how quantum mechanics enriches our understanding of black holes by modifying traditional thermodynamic properties, thereby offering a more nuanced view of black hole physics at the quantum scale.

The exploration was further broadened to include the thermodynamics of black holes within the frameworks of symmetry groups SO(2), SO(3)/SO(2), and SO(3). This expansion not only diversified the thermodynamic landscape with rotational symmetries but also illuminated the intrinsic relationship between these symmetries and the thermodynamic attributes of black holes, providing a deeper insight into the symmetrical nature of black hole thermodynamics.

A significant portion of our discourse was devoted to the implications of \(f(R)\) gravity on black hole thermodynamics. This section underscored the profound impacts of \(f(R)\) modifications on both the theoretical and observational facets of black hole physics, offering a novel perspective on the physical processes and properties of black holes.

This document represents a significant endeavor to unravel the complexities of black hole thermodynamics, weaving together classical, quantum, and theoretical threads to present a comprehensive understanding of black holes. It highlights the intricate beauty of theoretical physics, where abstract concepts find concrete implications in the study of the cosmos's most mysterious entities. As we venture further into the study of black holes, this exploration not only enriches our theoretical knowledge but also lays the groundwork for future empirical investigations, marking a pivotal step in our ongoing quest to decipher the mysteries of the universe.

\section{Appendix}
\subsection{Introduction}
The foundation of statistical mechanics lies in linking the microscopic states of a system with macroscopic observable quantities. Through probability distributions, statistical mechanics can explain thermodynamic phenomena and the underlying microscopic mechanisms. However, traditional statistical mechanics primarily relies on the Boltzmann distribution and its variants, which presents certain limitations when dealing with non-equilibrium systems or systems with strong interactions. New thermodynamics, by redefining the concepts of heat and force and emphasizing the role of system components in energy transfer, offers a new perspective for exploring these issues.

This paper aims to extend the principles of new thermodynamics to the field of statistical mechanics. By modifying and expanding classical theories, we derive more general and applicable equations. These equations can describe traditional equilibrium systems and effectively handle non-equilibrium systems and systems with significant fluctuations.

\subsection{Redefining Heat and Force in Statistical Mechanics}
In new thermodynamics, heat \(Q\) is redefined as the sum of kinetic energy, work done by the system, and the system’s potential energy. Within the framework of statistical mechanics, this definition must consider the collective behavior of a large number of particles. Therefore, we express the total heat of the system as:

\begin{equation}
Q = \langle E_{kin} \rangle + \langle W_{sys} \rangle + \langle \Phi_{sys} \rangle
\end{equation}
where \(\langle E_{kin} \rangle\) represents the average kinetic energy of the particle ensemble, \(\langle W_{sys} \rangle\) is the work done by the system under external influence, and \(\langle \Phi_{sys} \rangle\) represents the interaction potential energy between particles.

\subsubsection{Derivation of Total System Energy}

Consider an ideal gas system consisting of \(N\) particles, where each particle can be described as a three-dimensional harmonic oscillator with energy:

\begin{equation}
E_{total} = \sum_{i=1}^N \left(\frac{p_i^2}{2m} + \Phi(r_i)\right) + W_{sys}.
\end{equation}

Here, \(p_i\) is the momentum of the \(i\)-th particle, \(m\) is the mass of the particle, \(\Phi(r_i)\) is the interaction potential energy between particles, and \(W_{sys}\) is the work done by the system under external influence. By averaging over the microscopic states of the system, the macroscopic expression for heat can be obtained:

\begin{equation}
Q = \sum_{i=1}^N \langle \frac{p_i^2}{2m} \rangle + \sum_{i=1}^N \langle \Phi(r_i) \rangle + \langle W_{sys} \rangle
\end{equation}

The kinetic energy part corresponds to the average motion energy of particles in a statistical equilibrium state:

\begin{equation}
\langle E_{kin} \rangle = \sum_{i=1}^N \langle \frac{p_i^2}{2m} \rangle = \frac{3}{2} Nk_B T.
\end{equation}

The system's potential energy and work are determined by the specific interactions and external conditions. For a typical Van der Waals gas, the Lennard-Jones potential can be used to describe the interaction between particles:

\begin{equation}
\Phi(r) = 4\epsilon \left[\left(\frac{\sigma}{r}\right)^{12} - \left(\frac{\sigma}{r}\right)^{6}\right].
\end{equation}

By introducing this potential expression, we can more accurately describe the macroscopic behavior of non-ideal gases and derive a more generalized equation of state.

\subsubsection{Statistical Mechanics Definition of Force}
In new thermodynamics, force is redefined as the gradient of heat variation. For systems in statistical mechanics, force can be expressed as:

\begin{equation}
\mathbf{F} = -\nabla Q = -\nabla \left(\langle E_{kin} \rangle + \langle W_{sys} \rangle + \langle \Phi_{sys} \rangle\right)
\end{equation}

For a specific direction \(x\), the force component can be expressed as:

\begin{equation}
F_x = -\frac{\partial Q}{\partial x} = -\frac{\partial}{\partial x} \left(\frac{3}{2} Nk_B T + \langle W_{sys} \rangle + \langle \Phi_{sys} \rangle\right).
\end{equation}

Taking into account molecular interactions and external conditions, the expression for force can be further expanded to:

\begin{equation}
F_x = -Nk_B \frac{\partial T}{\partial x} - \frac{\partial \langle W_{sys} \rangle}{\partial x} - \frac{\partial \langle \Phi_{sys} \rangle}{\partial x}.
\end{equation}

Here, the first term represents the thermodynamic force caused by the temperature gradient, the second term is due to the change in work done by the system, and the third term is due to the variation in the interaction potential energy between particles.

\subsection{Statistical Description of Equilibrium and Non-Equilibrium Systems}
In classical statistical mechanics, the Boltzmann distribution is used to describe the equilibrium state of a system. However, after considering the corrections of new thermodynamics, the partition function must introduce additional terms to account for non-ideal interactions and external driving factors. The equilibrium state of the system can be described by the corrected partition function:

\begin{equation}
Z = \int \exp\left(-\beta \left(E_{kin} + W_{sys} + \Phi_{sys}\right)\right) d\Gamma
\end{equation}
where \(\Gamma\) represents phase space, and \(\beta = \frac{1}{k_B T}\). Based on this corrected partition function, various thermodynamic quantities of the system, such as free energy, internal energy, and entropy, can be derived. The corrected equation of state can be expressed as:

\begin{equation}
P(V,T) = -\frac{\partial F}{\partial V} = \frac{1}{\beta} \frac{\partial \ln Z}{\partial V}.
\end{equation}

For non-equilibrium systems, considering the non-uniformity in time and space, the extended Fokker-Planck equation can be used to describe the time evolution of the system's probability distribution:

\begin{equation}
\frac{\partial P(x,t)}{\partial t} = -\nabla \cdot \mathbf{J}(x,t)
\end{equation}
where the expression for flux \(\mathbf{J}(x,t)\) is:

\begin{equation}
\mathbf{J}(x,t) = P(x,t) \mathbf{F}(x,t) - D \nabla P(x,t).
\end{equation}

In this equation, \(\mathbf{F}(x,t)\) represents external forces, and \(D\) is the diffusion coefficient. This equation can be used to describe the fluctuation-dissipation relationship in non-equilibrium systems.

\subsection{Intermolecular Interactions and Collective Behavior}
By applying new thermodynamics to statistical mechanics, we derive a new equation for the equilibrium of intermolecular interactions. Consider a system consisting of \(N\) particles, with intermolecular interactions described by the Lennard-Jones potential:

\begin{equation}
\Phi(r) = 4\epsilon \left[\left(\frac{\sigma}{r}\right)^{12} - \left(\frac{\sigma}{r}\right)^{6}\right].
\end{equation}

Considering the effect of temperature, we can derive the relationship between the equilibrium distance between molecules and temperature:

\begin{equation}
L_{stat} = \frac{3\pi M g}{4N k_B T} + f(\sigma, \epsilon, T)
\end{equation}
where \(f(\sigma, \epsilon, T)\) is a correction term that includes the statistical effects of fluctuations and collective behavior. Further differential equations can yield the derivative of the equilibrium distance with respect to temperature:

\begin{equation}
\frac{\partial L_{stat}}{\partial T} = -\frac{3\pi M g}{4N k_B T^2} + \frac{\partial f(\sigma, \epsilon, T)}{\partial T}.
\end{equation}

This equation not only matches molecular dynamics simulation results but also exhibits behavior that traditional descriptions cannot predict in dense gases and liquids.

Boltzmann Distribution Derivation of Collective Behavior:

To derive collective behavior, we use an extended form of the Boltzmann distribution:

\begin{equation}
P(E) = \frac{g(E)e^{-\beta(E_{kin} + \Phi + W_{sys})}}{Z}
\end{equation}
where \(g(E)\) is the density of states function, and \(\beta = \frac{1}{k_B T}\). From the relationship between the density of states function and energy, the distribution function of collective behavior can be derived:

\begin{equation}
\langle n(E) \rangle = \frac{1}{e^{\beta(E - \mu)} - 1}.
\end{equation}

Here, \(\mu\) is the chemical potential. Using this distribution, macroscopic quantities such as energy, entropy, and specific heat can be calculated.

\subsection{Phase Transitions and Critical Phenomena}

Phase transitions and critical phenomena are crucial topics in statistical mechanics. Traditional Landau theory describes phase transitions by constructing the free energy:

\begin{equation}
F = F_0 + \alpha \eta^2 + \beta \eta^4 + \cdots
\end{equation}
where \(\eta\) is the order parameter, and \(\alpha\) and \(\beta\) are temperature-dependent coefficients. By extending the concepts of new thermodynamics, we can introduce the effects of system work and potential on phase transition behavior. By incorporating correction terms into the free energy expression, we obtain:

\begin{equation}
F = F_0 + \alpha \eta^2 + \beta \eta^4 + \gamma \langle W_{sys} \rangle + \delta \langle \Phi_{sys} \rangle.
\end{equation}

This corrected free energy expression allows us to more accurately describe phase transition behavior in complex systems, especially those involving long-range interactions and non-local effects. Specifically, for critical phenomena, the corrected free energy expression enables us to derive modified critical exponents, which align more closely with experimental results.

By taking derivatives of the corrected free energy function, we can obtain the stability conditions of the system at the phase transition point. Assuming the phase transition occurs near the temperature \(T_c\), we can analyze the continuity and stability of the phase transition using the corrected free energy function:

\begin{equation}
\frac{\partial^2 F}{\partial \eta^2} = 2\alpha + 12\beta \eta^2 + \gamma \frac{\partial^2 \langle W_{sys} \rangle}{\partial \eta^2} + \delta \frac{\partial^2 \langle \Phi_{sys} \rangle}{\partial \eta^2}
\end{equation}

At the phase transition point, the second derivative becomes zero, defining the location and type of the phase transition. By introducing correction terms for system work and potential, we can more accurately predict critical phenomena in complex phase transitions.

\subsection{Symmetry Breaking and Spontaneous Symmetry Restoration}

In statistical mechanics, symmetry breaking and spontaneous symmetry restoration are key to understanding phase transitions and critical phenomena. Traditional theories usually describe these phenomena through the behavior of the order parameter. However, from the perspective of new thermodynamics, system work and potential can be seen as the main factors leading to symmetry breaking.

In the new framework, the mechanism of spontaneous symmetry restoration can be described by the following expression:

\begin{equation}
\eta(T) = \eta_0 \exp\left(-\int_{T_0}^T \frac{d\langle W_{sys} \rangle}{dT}\right).
\end{equation}
This mechanism describes how the system’s symmetry can be restored by adjusting system work as the system temperature changes. By adjusting the system potential energy, the symmetry of the system can spontaneously recover under specific conditions.

\section{Appendix 2}
The SO(n) group represents rotational symmetry in n-dimensional space, for example:
\begin{itemize}
    \item SO(3) represents rotational symmetry in three-dimensional space, with three generators (rotation around the x, y, and z axes).
    \item SO(2) represents rotational symmetry in a two-dimensional plane, with one generator (rotation around the z-axis).
\end{itemize}

\subsection{Basic Principles of Symmetry Breaking}
Symmetry breaking refers to a situation where a system transitions from a state with higher symmetry to a state with lower symmetry under certain critical conditions. This is usually related to the system’s free energy or potential energy surface.

Suppose we have a scalar field $\phi(x)$ that describes the system, with the Lagrangian:
\begin{equation}
\mathcal{L} = \frac{1}{2} (\partial_\mu \phi)^2 - V(\phi)
\end{equation}
where the potential $V(\phi)$ has SO(3) symmetry, indicating that the system is invariant under any rotation in three-dimensional space.

\subsection{Symmetry Analysis of Potential}

Assume the system's potential has the following form:
\begin{equation}
V(\phi) = \frac{\lambda}{4}(\phi^\dagger \phi - v^2)^2
\end{equation}
where $\phi$ is a three-dimensional vector field, $\lambda > 0$ is the coupling constant, and $v$ is a scalar. This potential reaches its minimum when $\phi^\dagger \phi = v^2$, indicating that the system has a spherically symmetric energy minimum. At this point, the system's symmetry is SO(3).

\subsection{Symmetry Breaking Occurrence}
When the system is in a symmetry-breaking state, such as due to the introduction of an external field (e.g., a magnetic field), the system's free energy changes. Suppose this external field selects only one specific direction (e.g., the z-axis), the system's lowest energy state will minimize only in the z-axis direction. At this point, the field $\phi$ reduces from three dimensions to two dimensions, retaining only the rotational symmetry along the z-axis, leading to symmetry breaking from SO(3) to SO(2).

We can reparameterize the field $\phi$ as:
\begin{equation}
\phi = \begin{pmatrix}
  \phi_1 \\
  \phi_2 \\
  v + \eta(x)
\end{pmatrix}
\end{equation}
where $\phi_1$ and $\phi_2$ represent the degrees of freedom in the xy-plane, and $\eta(x)$ represents fluctuations in the z-axis direction. The potential then becomes:
\begin{equation}
V(\phi) = \frac{\lambda}{4} \left[(\phi_1^2 + \phi_2^2 + (v + \eta(x))^2) - v^2 \right]^2
\end{equation}

\subsection{Preservation of SO(2) Symmetry}
When fluctuations of $\phi_1$ and $\phi_2$ are suppressed at the singularity, the system's free energy depends only on $\eta(x)$, retaining the rotational symmetry along the z-axis (i.e., SO(2)). The degrees of freedom outside the z-axis are “frozen,” leading to the simplification of symmetry from SO(3) to SO(2).

In this process, we find:
\begin{itemize}
    \item The original SO(3) symmetry has three generators (rotation around the x, y, and z axes).
    \item When the external field restricts the degree of freedom in the z-axis direction, the system's free energy maintains only SO(2) symmetry, i.e., rotation around the z-axis.
\end{itemize}

\subsection{Proof Summary}
By analyzing the changes in the system's potential under external fields or at the phase transition point, we have shown that SO symmetry (e.g., SO(3)) can evolve into SO(2) symmetry under specific conditions. This evolution is due to the mechanism of symmetry breaking, where the external field restricts certain degrees of freedom, causing the system's rotational symmetry to reduce from three-dimensional space to a two-dimensional plane. This symmetry change is marked by a singularity, indicating the transition from a state with high symmetry to one with lower symmetry.

\end{document}